\RequirePackage{lineno}
\documentclass[aps,prl,twocolumn,showpacs,superscriptaddress,groupedaddress]{revtex4}  

 \usepackage{graphicx}

\usepackage{dcolumn}   
\usepackage{bm}        
\usepackage{amssymb}   

\newcommand{\met}{\mbox{\ensuremath{\not \!\! E_T}}}
\newcommand{\vmet}{\mbox{\ensuremath{\not \!\! \vec{E}_T}}}

\usepackage{xspace}
\newcommand{\dzero}{D0\xspace}

\begin{document}
\hspace{5.2in} \mbox{Fermilab-Pub-11/663-E}

\title{Search for universal extra dimensions in $\boldsymbol{p\bar p}$ collisions}

%
\affiliation{Universidad de Buenos Aires, Buenos Aires, Argentina}
\affiliation{LAFEX, Centro Brasileiro de Pesquisas F{\'\i}sicas, Rio de Janeiro, Brazil}
\affiliation{Universidade do Estado do Rio de Janeiro, Rio de Janeiro, Brazil}
\affiliation{Universidade Federal do ABC, Santo Andr\'e, Brazil}
\affiliation{Instituto de F\'{\i}sica Te\'orica, Universidade Estadual Paulista, S\~ao Paulo, Brazil}
\affiliation{University of Science and Technology of China, Hefei, People's Republic of China}
\affiliation{Universidad de los Andes, Bogot\'{a}, Colombia}
\affiliation{Charles University, Faculty of Mathematics and Physics, Center for Particle Physics, Prague, Czech Republic}
\affiliation{Czech Technical University in Prague, Prague, Czech Republic}
\affiliation{Center for Particle Physics, Institute of Physics, Academy of Sciences of the Czech Republic, Prague, Czech Republic}
\affiliation{Universidad San Francisco de Quito, Quito, Ecuador}
\affiliation{LPC, Universit\'e Blaise Pascal, CNRS/IN2P3, Clermont, France}
\affiliation{LPSC, Universit\'e Joseph Fourier Grenoble 1, CNRS/IN2P3, Institut National Polytechnique de Grenoble, Grenoble, France}
\affiliation{CPPM, Aix-Marseille Universit\'e, CNRS/IN2P3, Marseille, France}
\affiliation{LAL, Universit\'e Paris-Sud, CNRS/IN2P3, Orsay, France}
\affiliation{LPNHE, Universit\'es Paris VI and VII, CNRS/IN2P3, Paris, France}
\affiliation{CEA, Irfu, SPP, Saclay, France}
\affiliation{IPHC, Universit\'e de Strasbourg, CNRS/IN2P3, Strasbourg, France}
\affiliation{IPNL, Universit\'e Lyon 1, CNRS/IN2P3, Villeurbanne, France and Universit\'e de Lyon, Lyon, France}
\affiliation{III. Physikalisches Institut A, RWTH Aachen University, Aachen, Germany}
\affiliation{Physikalisches Institut, Universit{\"a}t Freiburg, Freiburg, Germany}
\affiliation{II. Physikalisches Institut, Georg-August-Universit{\"a}t G\"ottingen, G\"ottingen, Germany}
\affiliation{Institut f{\"u}r Physik, Universit{\"a}t Mainz, Mainz, Germany}
\affiliation{Ludwig-Maximilians-Universit{\"a}t M{\"u}nchen, M{\"u}nchen, Germany}
\affiliation{Fachbereich Physik, Bergische Universit{\"a}t Wuppertal, Wuppertal, Germany}
\affiliation{Panjab University, Chandigarh, India}
\affiliation{Delhi University, Delhi, India}
\affiliation{Tata Institute of Fundamental Research, Mumbai, India}
\affiliation{University College Dublin, Dublin, Ireland}
\affiliation{Korea Detector Laboratory, Korea University, Seoul, Korea}
\affiliation{CINVESTAV, Mexico City, Mexico}
\affiliation{Nikhef, Science Park, Amsterdam, the Netherlands}
\affiliation{Radboud University Nijmegen, Nijmegen, the Netherlands and Nikhef, Science Park, Amsterdam, the Netherlands}
\affiliation{Joint Institute for Nuclear Research, Dubna, Russia}
\affiliation{Institute for Theoretical and Experimental Physics, Moscow, Russia}
\affiliation{Moscow State University, Moscow, Russia}
\affiliation{Institute for High Energy Physics, Protvino, Russia}
\affiliation{Petersburg Nuclear Physics Institute, St. Petersburg, Russia}
\affiliation{Instituci\'{o} Catalana de Recerca i Estudis Avan\c{c}ats (ICREA) and Institut de F\'{i}sica d'Altes Energies (IFAE), Barcelona, Spain}
\affiliation{Stockholm University, Stockholm and Uppsala University, Uppsala, Sweden}
\affiliation{Lancaster University, Lancaster LA1 4YB, United Kingdom}
\affiliation{Imperial College London, London SW7 2AZ, United Kingdom}
\affiliation{The University of Manchester, Manchester M13 9PL, United Kingdom}
\affiliation{University of Arizona, Tucson, Arizona 85721, USA}
\affiliation{University of California Riverside, Riverside, California 92521, USA}
\affiliation{Florida State University, Tallahassee, Florida 32306, USA}
\affiliation{Fermi National Accelerator Laboratory, Batavia, Illinois 60510, USA}
\affiliation{University of Illinois at Chicago, Chicago, Illinois 60607, USA}
\affiliation{Northern Illinois University, DeKalb, Illinois 60115, USA}
\affiliation{Northwestern University, Evanston, Illinois 60208, USA}
\affiliation{Indiana University, Bloomington, Indiana 47405, USA}
\affiliation{Purdue University Calumet, Hammond, Indiana 46323, USA}
\affiliation{University of Notre Dame, Notre Dame, Indiana 46556, USA}
\affiliation{Iowa State University, Ames, Iowa 50011, USA}
\affiliation{University of Kansas, Lawrence, Kansas 66045, USA}
\affiliation{Kansas State University, Manhattan, Kansas 66506, USA}
\affiliation{Louisiana Tech University, Ruston, Louisiana 71272, USA}
\affiliation{Boston University, Boston, Massachusetts 02215, USA}
\affiliation{Northeastern University, Boston, Massachusetts 02115, USA}
\affiliation{University of Michigan, Ann Arbor, Michigan 48109, USA}
\affiliation{Michigan State University, East Lansing, Michigan 48824, USA}
\affiliation{University of Mississippi, University, Mississippi 38677, USA}
\affiliation{University of Nebraska, Lincoln, Nebraska 68588, USA}
\affiliation{Rutgers University, Piscataway, New Jersey 08855, USA}
\affiliation{Princeton University, Princeton, New Jersey 08544, USA}
\affiliation{State University of New York, Buffalo, New York 14260, USA}
\affiliation{Columbia University, New York, New York 10027, USA}
\affiliation{University of Rochester, Rochester, New York 14627, USA}
\affiliation{State University of New York, Stony Brook, New York 11794, USA}
\affiliation{Brookhaven National Laboratory, Upton, New York 11973, USA}
\affiliation{Langston University, Langston, Oklahoma 73050, USA}
\affiliation{University of Oklahoma, Norman, Oklahoma 73019, USA}
\affiliation{Oklahoma State University, Stillwater, Oklahoma 74078, USA}
\affiliation{Brown University, Providence, Rhode Island 02912, USA}
\affiliation{University of Texas, Arlington, Texas 76019, USA}
\affiliation{Southern Methodist University, Dallas, Texas 75275, USA}
\affiliation{Rice University, Houston, Texas 77005, USA}
\affiliation{University of Virginia, Charlottesville, Virginia 22901, USA}
\affiliation{University of Washington, Seattle, Washington 98195, USA}
\author{V.M.~Abazov} \affiliation{Joint Institute for Nuclear Research, Dubna, Russia}
\author{B.~Abbott} \affiliation{University of Oklahoma, Norman, Oklahoma 73019, USA}
\author{B.S.~Acharya} \affiliation{Tata Institute of Fundamental Research, Mumbai, India}
\author{M.~Adams} \affiliation{University of Illinois at Chicago, Chicago, Illinois 60607, USA}
\author{T.~Adams} \affiliation{Florida State University, Tallahassee, Florida 32306, USA}
\author{G.D.~Alexeev} \affiliation{Joint Institute for Nuclear Research, Dubna, Russia}
\author{G.~Alkhazov} \affiliation{Petersburg Nuclear Physics Institute, St. Petersburg, Russia}
\author{A.~Alton$^{a}$} \affiliation{University of Michigan, Ann Arbor, Michigan 48109, USA}
\author{G.~Alverson} \affiliation{Northeastern University, Boston, Massachusetts 02115, USA}
\author{M.~Aoki} \affiliation{Fermi National Accelerator Laboratory, Batavia, Illinois 60510, USA}
% ...
%
%
\collaboration{The D0 Collaboration\footnote{with visitors from
$^{a}$Augustana College, Sioux Falls, SD, USA,
$^{b}$The University of Liverpool, Liverpool, UK,
$^{c}$UPIITA-IPN, Mexico City, Mexico,
$^{d}$SLAC, Menlo Park, CA, USA,
$^{e}$University College London, London, UK,
$^{f}$Centro de Investigacion en Computacion - IPN, Mexico City, Mexico,
$^{g}$ECFM, Universidad Autonoma de Sinaloa, Culiac\'an, Mexico,
and 
$^{h}$Universit{\"a}t Bern, Bern, Switzerland,
$^{i}$DESY, Hamburg, Germany.
$^{\ddag}$Deceased.
}} \noaffiliation
\vskip 0.25cm
      
\date{February 3, 2012}

\begin{abstract} We present a search for Kaluza-Klein (KK) particles predicted by models with universal extra dimensions (UED)  
using a data set  corresponding to an integrated
luminosity of $7.3$~fb$^{-1}$, collected by the \dzero detector at a $p\bar p$
center of mass energy of 1.96 TeV. The decay chain of KK particles can lead to a final state with two muons of the same charge.
This signature is used to set a
lower limit on the compactification scale of $R^{-1}>260$~GeV in a minimal UED model.
\end{abstract}

\maketitle

The existence of extra dimensions in addition to the $3+1$ dimensions of space-time 
has been postulated as a possible solution to the problem of the large hierarchy of scales
in the standard model (SM).  In models with
universal extra dimensions (UED) all particles propagate in the extra dimensions~\cite{ued1}.
In this Letter, we study a minimal UED (mUED) model, which has only one extra dimension~\cite{ued2}.

Each SM particle in the mUED model 
is associated with a set of excited Kaluza-Klein (KK) states when viewed in $3+1$ dimensions.
Since compactification of the extra dimensions leads to periodic boundary conditions, 
the KK states have discrete masses of the order of $R^{-1}$, where
$R$ is the radius of the compact dimension. 
If one-loop corrections are applied, the mass spectrum of the KK modes also depends on a cutoff scale 
for boundary terms, which is chosen to be 10~TeV~\cite{ued2}.
Gluon KK modes ($g_1$) are the heaviest particles,
followed by quarks (SU(2) doublet $Q_1$ or singlet $q_1$), gauge bosons
($Z_1/W_1$), leptons (SU(2) doublet $L_{1}$ or singlet $\ell_1$) and the KK photon ($\gamma_1$), which
is the lightest KK particle (LKP) and does not decay. The LKP is also a dark matter candidate~\cite{LKP}.

Previous searches for KK particles predicted by a modified UED model have been 
conducted by the \dzero~\cite{dzeroDiphoton} Collaboration
at the Fermilab Tevatron Collider and by the ATLAS~\cite{atlasDiphoton} Collaboration at the CERN
Large Hadron Collider (LHC) in diphoton final states. In this model, gravity mediation allows
decays of the LKP into a photon plus a light KK graviton.
A search for events with two leptons of the same
charge, which can be interpreted in a mUED model, has been 
performed by ATLAS~\cite{atlasDilepton}, but 
no dedicated study of the mUED model has so far been performed at a collider.
Constraints from precision electroweak data~\cite{ued1}, from the
measurements of the muon anomalous magnetic moment and
of the $b\to s \gamma$ branching ratio~\cite{ued3}, indicate that the scale
$R^{-1}$ can be as low as $\approx 300$ GeV, making KK
particles accessible at the Tevatron. In this Letter, the
first search for mUED in the final state with two leptons
of the same charge is presented using data
corresponding to 7.3~fb$^{-1}$ of integrated luminosity collected by the \dzero
detector~\cite{dzero} at $\sqrt{s}=1.96$~TeV.

\begin{figure}[!htb]
\centering
\includegraphics[scale=0.6]{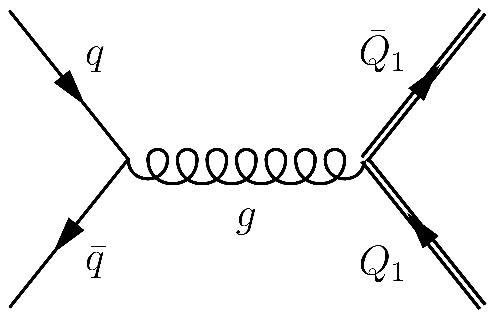} 
\includegraphics[scale=0.6]{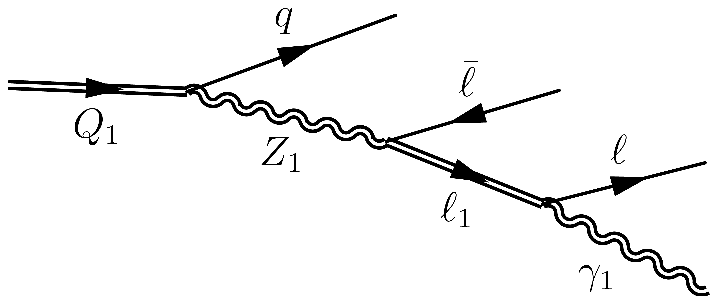} 
\caption{(a) Production of a pair of KK quarks ($Q_1\bar Q_1$).
(b) Decay of a KK quark into a jet, two oppositely charged leptons, and the LKP. 
Double lines indicate KK excitations.
A similar cascade decay occurs for the second KK quark leading 
to several leptons of the same charge in the final state.}
\label{fig:feynman}
\end{figure}

At the Tevatron, KK gluons or quarks are mainly produced in pairs, 
as shown in Fig.~\ref{fig:feynman}.
In the subsequent cascade decay, up to four charged leptons are produced. 
Since the masses of the extra particles predicted by the mUED models are nearly degenerate,
the leptons are emitted with low transverse momentum and might escape detection.
In this analysis, we select events with two muons of the same charge.

The \dzero detector~\cite{dzero} consists of tracking systems and calorimeters.
The innermost part is a tracking system where
charged particles are detected by the silicon microstrip (SMT)
and central fiber (CFT) tracking detectors, located within a 2~T solenoid. 
The tracking system is surrounded by a
liquid-argon/uranium calorimeter. Particle energies are measured in the
electromagnetic and hadronic calorimeters within a pseudorapidity range of
$|\eta|<4.2$~\cite{pseudorapidities}.
Jets are reconstructed with a cone algorithm using a radius of ${\cal R}=0.5$~\cite{jetCone} in the
calorimeter. The central and forward muon detectors are
composed of a layer of wire chambers and scintillators in front of a 1.8~T
toroid magnet and two layers outside the toroid. Missing transverse energy 
is measured from the vector sum of the calorimeter cell energies in the $xy$ plane~\cite{missinget}. 
A correction for
the energy response of muons, electrons and jets is applied.

The backgrounds from $Z$+jets, $W$+jets, and $t\bar{t}$ production are modeled by the {\sc alpgen}~\cite{alpgen}
Monte Carlo (MC) event generator, interfaced with {\sc pythia}~\cite{pythia6} for
showering and hadronization.
Diboson production ({\it WW}, {\it WZ} and {\it ZZ}) is simulated by {\sc pythia}. 
The CTEQ6L1 parametrization of the parton 
distribution functions (PDF) are used~\cite{CTEQ6L1}.
Higher order cross sections for diboson and $W/Z$+jets production are calculated by {\sc mcfm}~\cite{MCFM}
and the cross section of $t\bar t$ pair production is taken from~\cite{ttCrossSection}.
The dominant source of background is pairs of muons with the same charge from heavy flavor jets.
This background contribution is estimated from data and is described in detail in the following. 

Signal MC events are generated for 9 different values of $R^{-1}$, covering
the range from 200 to 320 GeV in steps of 15 GeV, using 
{\sc pythia} with the CTEQ5L PDF parametrization~\cite{CTEQ5}.
The production cross sections and masses of KK particles are taken from {\sc pythia}. They
are given in Table~\ref{tab:UED_Xsections_KK_masses} for each $R^{-1}$ and with all KK gluon and quark production modes included. 
All decay mechanisms leading to like-charge dimuon final states are taken into account. The decay branching fractions for all KK particles are given by the mUED model.
After simulating all cascade decays, approximately $1\%$ of events have two like-charged muons.

\begin{table}[!hbt]
\begin{ruledtabular}
\caption{\label{tab:UED_Xsections_KK_masses} Masses of KK particles 
for each $R^{-1}$ value used in the MC generation 
with corresponding total production cross section.}
\begin{tabular}{ccccccc} 
$R^{-1}$  &  \multicolumn{5}{c}{Masses (GeV)} & Cross Section \\ 
(GeV)  &    $\gamma_1$    &   $Z_1$ &   $g_1$ &   $\ell_1$ &   $Q_1$	&  (pb) \\  \hline
200     &   201     &   230 &   269 &   207 &   249 &	$34.9	\pm 0.2$ \\
215     &   216     &   245 &   287 &   222 &   266 &	$20.4	\pm 0.1$ \\
230     &   231     &   260 &   305 &   238 &   283 &	$12.1	\pm 0.1$ \\
245     &   246     &   274 &   323 &   253 &   300 &	$7.24	\pm 0.05$ \\
260     &   261     &   289 &   341 &   268 &   317 &	$4.39	\pm 0.03$ \\
275     &   276     &   304 &   359 &   284 &   334 &	$2.69	\pm 0.02$ \\
290     &   291     &   319 &   377 &   299 &   351 &	$1.65	\pm 0.01$ \\
305     &   306     &   335 &   395 &   314 &   368 &	$1.02	\pm 0.06$ \\
320     &   321     &   350 &   413 &   330 &   385 &	$0.63	\pm 0.01$ \\
\end{tabular}
\end{ruledtabular}
\end{table}

All signal and background MC events pass through the full
{\sc geant}-based simulation of the detector~\cite{geant} and are reconstructed using the same algorithms as used for data.
To simulate detector noise and multiple $p\bar p$ interactions, MC events are overlaid with data
events from random beam crossings.

Events are selected by requiring that they pass at least one single muon trigger condition. 
We require that each event must contain at least two muons of the same charge. The track in the muon system must be matched
to a track in the central tracking system with detector $|\eta|<1.5$. We reject cosmic rays by requiring
the associated scintillator hits in the muon system to be consistent with originating from a $p\bar p$ collision,
the distance of closest approach (dca) of the muon tracks to the $p\bar p$ interaction vertex to be less than $0.05$~cm,
and the differences between the $z$ coordinates of the dca of each muon and the $p\bar p$ interaction vertex to be $<1$~cm. 
We require  $\Delta\phi_{\mu\mu}<2.9$~rad for the azimuthal angle between the two muons in order
to reject multijet background in which there are back-to-back jets which can contain muons from semileptonic meson decay, and
$\Delta\phi_{\mu\mu}>0.25$~rad to reject muons originating from the same jet. 
The transverse momenta of the leading and next-to-leading muons have to be $15<p_{T_1}<200$~GeV and $p_{T_2}>10$~GeV, respectively.
The invariant mass of the muon pair must be in the range $M_{\mu\mu}<250$~GeV. The upper limits on the muon $p_T$ and dimuon invariant mass reduce the number of events with one of the muon charges mis-reconstructed. In addition,
we reject events with $\met<25$~GeV, since multijet background dominates at low $\met$. 

To discriminate between isolated muons from
signal and muons contained in jets, we define the isolation in the calorimeter, ${\cal I}^{cal}$,
 as the sum of the energy deposited in the calorimeter
within an annulus of $0.1<{\cal R}<0.4$, divided by the muon $p_T$, and the isolation in the tracking detector,
${\cal I}^{trk}$, as the sum of the $p_T$ of all charged particles
within a cone of ${\cal R}=0.5$ around the muon, excluding the muon itself, divided by $p_T^{\mu}$. 
At least one of the two muons is required to be 
isolated with ${\cal I}^{cal} < 0.4$ and ${\cal I}^{trk}<0.12$. 

We estimate the multijet background by defining a signal and a background enriched sample.
The signal enriched sample comprises events where the requirement on the isolation
of the second muon in the tracking detector is relaxed to ${\cal I}^{trk}<0.25$.  
To define a background
enriched sample, we require that the second muon fails the isolation requirement.
A normalization factor is calculated for each jet multiplicity, given by the ratio of the number of events in the signal
and background enriched samples in a multijet dominated region where the $p_T$ of the most isolated muon is $5<p_T<10$~GeV.
The multijet background is determined by multiplying
the background enriched sample with this normalisation factor in the region where the $p_T$ of the most isolated muon is $>10$~GeV.

In the high $p_T$ region, there is a significant contribution to the background enriched sample from
SM processes other than multijet,
particularly from $W$+jets production, with an isolated muon from the $W$ boson 
decay and a non-isolated muon embedded in the jet.
This SM background is modeled by applying the selection used to define the background enriched sample to MC events.
The resulting SM background distributions, corrected with the same normalization factors as used for data,
are subtracted from the corrected distribution of the background enriched data sample 
to obtain an estimate of the multijet background. 

Events from $Z\rightarrow \mu\mu$ decays enter our sample mainly through mis-reconstruction of one of the muon charges.
This background is estimated from simulation. We 
use data to determine the corresponding systematic uncertainty by comparing the
two independent charge measurements in the central tracking detector and the muon detector~\cite{chargeflip}. 
We determine the rate of muons with an incorrectly measured charge by counting the number of events
where the two measurements disagree.  The number of charge-flip events in data 
is in good agreement with the number of like-charged $Z \rightarrow \mu\mu$ events predicted by the simulation.

\begin{figure}[!hbt]
  \centering
  \includegraphics[width=7cm]{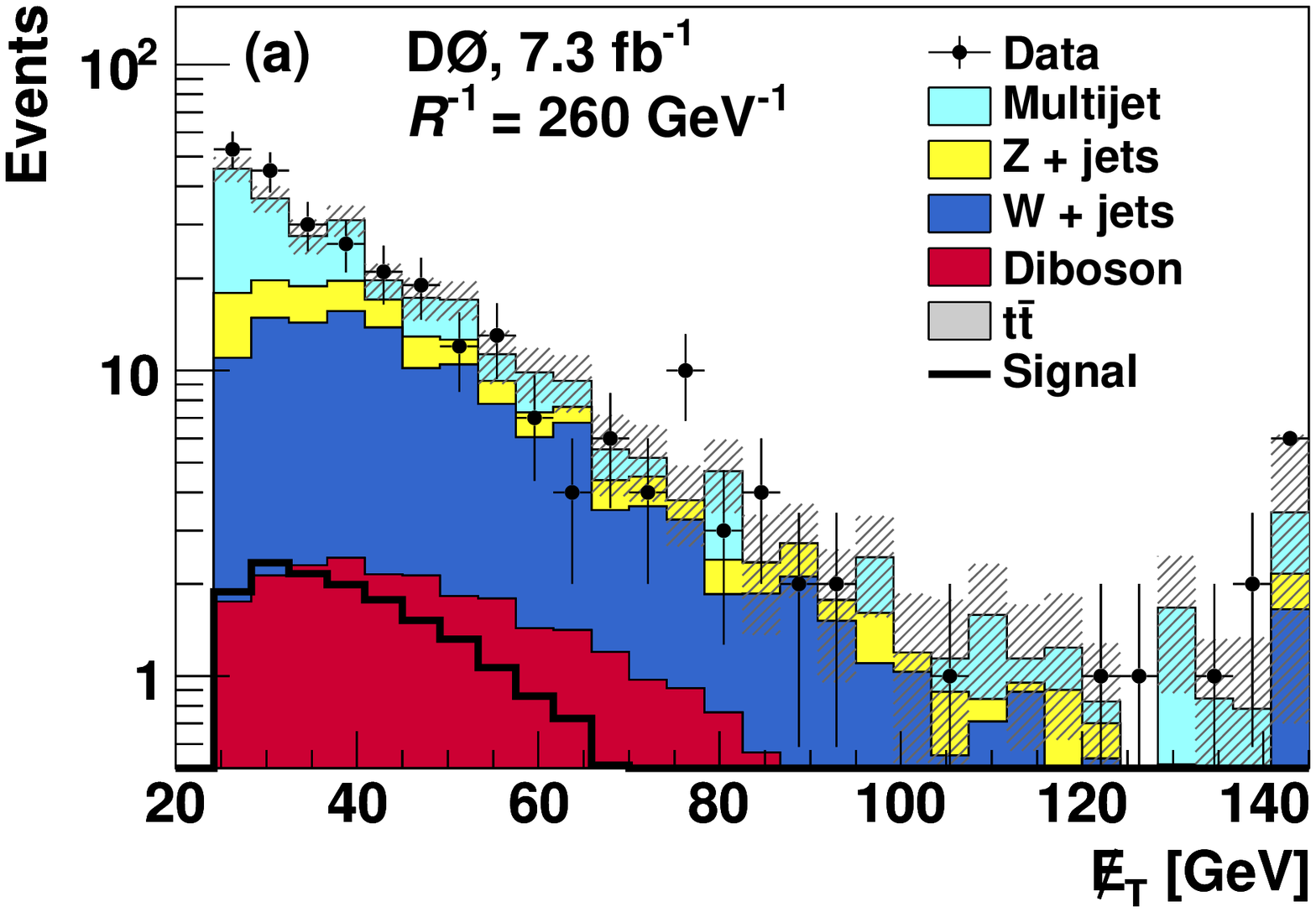}\\
  \includegraphics[width=7cm]{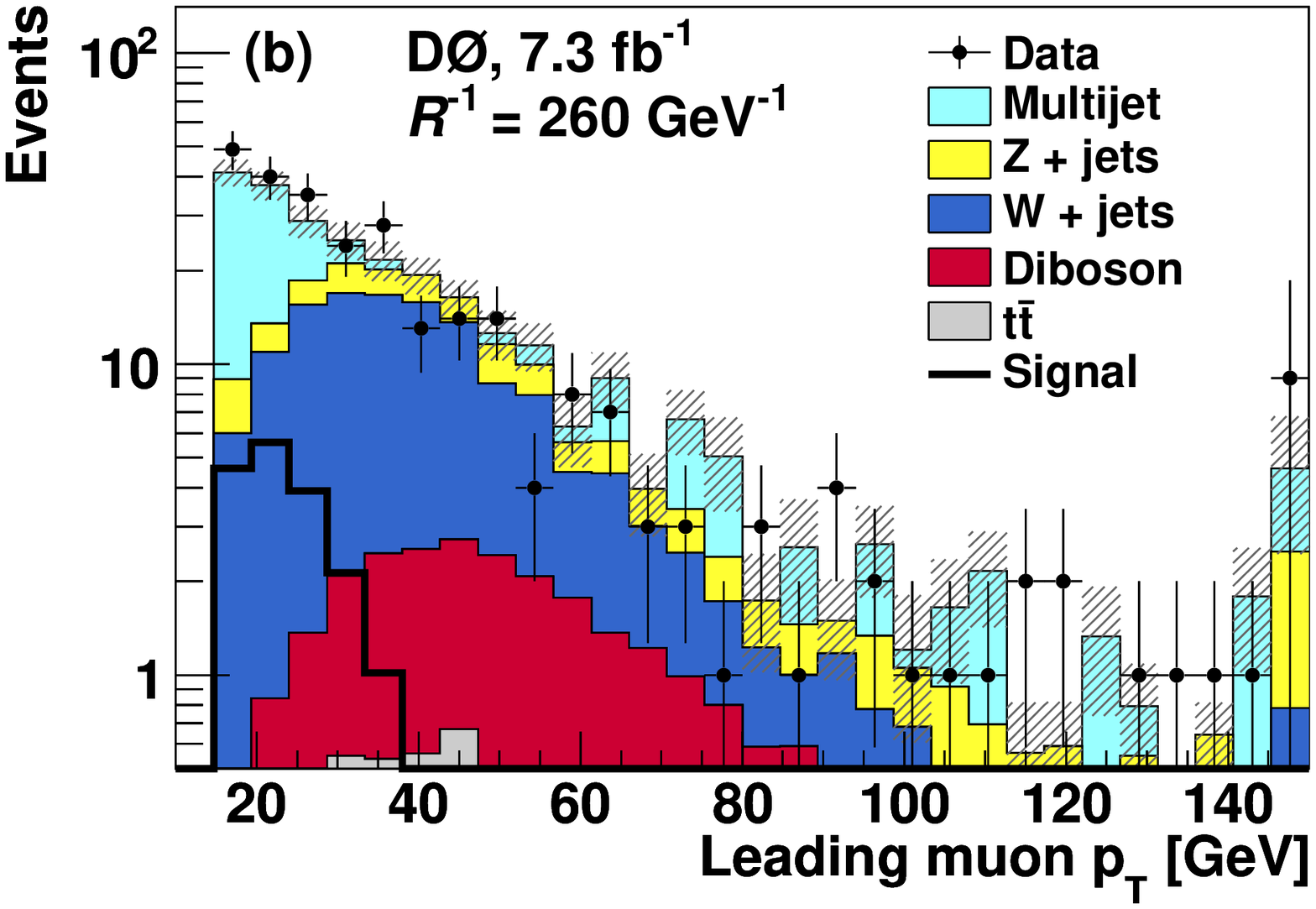}  
    \caption{
  Distribution of (a) $\met$ and (b) $p_{T_1}$ for data and background, compared
   to a signal with $R^{-1}=260$~GeV.
  The shaded band shows the statistical uncertainty on the background estimation.
  All entries exceeding the range of the
     histogram are added to the last bin.}
\label{fig:input_to_TMVA}
\end{figure}

To improve the discrimination between signal and background
a boosted decision tree (BDT)~\cite{TMVA} is trained using 
$p_{T_1}$, $p_{T_2}$, $M_{\mu\mu}$, $\Delta\phi_{\mu\mu}$, and the number of jets
as input variables. Further input variables are the
$\chi^2$ of the fit of the muon tracks to reject badly reconstructed muons, the transverse masses
$M_{T} = \sqrt{2 \met p_{T} (1-\cos\Delta\phi(\vmet,\mu))}$ calculated separately for each muon, and the scalar 
product of $\met$ and $p_{T_2}$. This rejects multijet background events with small $\met$ or $p_{T_2}$.

\begin{figure}[!hbt]
  \centering
  \includegraphics[width=7cm]{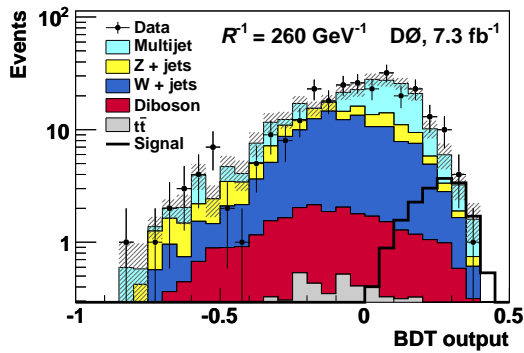}
  \caption{Distribution of the BDT output for data and background, compared
   to a signal with $R^{-1}=260$~GeV.
  The shaded band shows the statistical uncertainty on the background estimation.}
  \label{fig:tmva_output}
\end{figure}

The distributions of $\met$ and $p_{T_1}$ for data and background are shown Fig.~\ref{fig:input_to_TMVA},
together with signal for $R^{-1}=260$~GeV. 
The BDT is trained separately for each signal point with different $R^{-1}$. The BDT output distribution
is shown in Fig.~\ref{fig:tmva_output} and the event yields in
data are compared to the expected number of background and signal events in Table~\ref{tab-events}.
Signal is concentrated in the region where the BDT output is $>0$.
 
\begin{table}
\caption{Expected number of events for backgrounds, event yields in data and 
expected number of events for a signal with $R^{-1}=260$~GeV after the final selection and requiring 
a BDT output $>0$ (for illustrative purpose, this cut is not used for limit estimation). The total uncertainties are also given.
 \label{tab-events}}
\begin{tabular}{lcc}
\hline\hline
Process &  Final Selection &  BDT output $>0$  \\
\hline
Diboson    	    &   $21 \pm 3$    & $\phantom{0}6 \pm 1$ \\
$Z$+jets     	    &   $39 \pm 9$    & $13 \pm 3$\\
$W$+jets     	    &   $109\pm 14$ & $38 \pm 5$\\
$t\bar{t}$               &   $\phantom{0}6 \pm 1$     & $\phantom{0}2\pm 1$\\
Multijet                  &   $\phantom{0}95 \pm 41$ & $\phantom{0}63 \pm 27$ \\
Total Background  &   $271 \pm 45$  & $123 \pm  28$ \\ \hline
Data                      &   $273$            & $126$   \\  \hline
Signal                    &  $18 \pm 1$  & $18 \pm 1$ \\
\hline
\hline
\end{tabular}
\end{table}

Systematic uncertainties on the normalisation of both background and signal, 
including their correlations, are taken into account. These include
theoretical uncertainties on SM background cross sections ($7\%-15\%$) and
the uncertainties on the rate of multijet background ($40\%$), signal efficiency from
choice of PDF parametrization ($4\%$), integrated
luminosity ($6.1\%$), jet energy scale ($4\%$), rate of $Z/\gamma^*$ production with
an incorrectly reconstructed muon charge ($21\%$), trigger efficiency ($6\%$) and 
muon reconstruction and isolation ($2\%$).

\begin{figure}[!hbt]
\includegraphics[width=7cm]{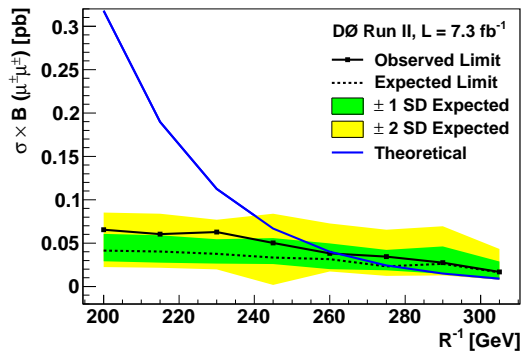}
\caption{\label{fig:limits1} Observed and median expected  $95\%$~CL
limits on $\sigma{\cal B} (\mu^{\pm}\mu^{\pm})$ as a function of $R^{-1}$
compared to $\sigma {\cal B} (\mu^{\pm}\mu^{\pm})$ calculated with the mUED model.
The bands represent $\pm 1$ and $ \pm 2$ standard deviations around the
median expected limits.}
\end{figure}

The entire differential distribution of the BDT output is used to set
limits on the product of the cross section and the branching ratio with the
{\sl CL}$_s$ method~\cite{bib-junk} and a profiling technique to reduce the impact of 
systematic uncertainties~\cite{collie}.
Observed and median expected upper limits at $95\%$ CL on the product of
the cross section $\sigma$ and the branching fraction ${\cal B}$ into like-charged muon pairs 
are shown  in Fig.~\ref{fig:limits1} as a function of $R^{-1}$.
The theoretical cross section of the mUED model intersects the expected limit at
$R^{-1}=275$~GeV and the observed at $R^{-1}=260$~GeV, 
which corresponds to a mass of $317$~GeV of the lightest
KK quark in the mUED model.

In summary, we present a search for extra dimensions in the mUED model using 7.3~fb$^{-1}$
of integrated luminosity collected by the \dzero experiment. The first direct 
lower limit on the compactification scale of the extra dimension in the mUED model
is set as $R^{-1}=260$~GeV.

%
We thank the staffs at Fermilab and collaborating institutions,
and acknowledge support from the
DOE and NSF (USA);
CEA and CNRS/IN2P3 (France);
FASI, Rosatom and RFBR (Russia);
CNPq, FAPERJ, FAPESP and FUNDUNESP (Brazil);
DAE and DST (India);
Colciencias (Colombia);
CONACyT (Mexico);
NRF (Korea);
CONICET and UBACyT (Argentina);
FOM (The Netherlands);
STFC and the Royal Society (United Kingdom);
MSMT and GACR (Czech Republic);
BMBF and DFG (Germany);
SFI (Ireland);
The Swedish Research Council (Sweden);
and
CAS and CNSF (China).


\begin{thebibliography}{99}

\bibitem{ued1}
T.~Appelquist, H.-C.\ Cheng and B.\ A.\ Dobrescu, Phys.\ Rev.\ D {\bf 64}, 035002 (2001).

\bibitem{ued2}
H.-C.\ Cheng, K.~T.~Matchev and M.~Schmaltz, Phys.\ Rev.\ D {\bf 66}, 056006 (2002).

\bibitem{LKP}
H.-C.\ Cheng, J.~L.~Feng and K.~T.~Matchev, Phys.\ Rev.\ Lett. {\bf 89}, 211301 (2002).

\bibitem{dzeroDiphoton}
V.~M.~Abazov {\sl et al.}, (D0 Collaboration), Phys. Rev. Lett. {\bf 105}, 221802 (2010).

\bibitem{atlasDiphoton}
G.~Aad {\it et al.} (ATLAS Collaboration),  Eur.\ Phys.\ J.\ C {\bf 71}, 1744 (2011).

\bibitem{atlasDilepton}
G.~Aad {\it et al.} (ATLAS Collaboration), J.\ High Energy Phys.\ {\bf 10}, 107 (2011).

\bibitem{ued3}
K.~Agashe, N.~G.~Deshpande and G.~H.~Wu, Phys.\ Lett.\ B {\bf 514}, 309 (2001);
T.~Appelquist and B.~A.~Dobrescu, Phys.\ Lett.\ B {\bf 516}, 85 (2001).

\bibitem{dzero}
V.~M.~Abazov {\sl et al}. (D0 Collaboration), Nucl.\ Instrum.\ Methods Phys.\ Res., Sect.\ A {\bf 565}, 463 (2006).

\bibitem{pseudorapidities}
The \dzero detector coordinate system is right-handed with the $z$ axis pointing in the proton beam direction.
The $y$ axis points upward and the azimuthal angle $\phi$ is measured from the $x$ axis. Pseudorapidity is defined as $\eta = -\ln[\tan(\theta/2)]$, where $\theta$ is the polar angle.

\bibitem{jetCone}
G.~C.~Blazey {\sl et al.}, arXiv:hep-ex/0005012 (2000). 

\bibitem{missinget}
Missing transverse energy is defined as $\vmet=-\displaystyle\sum_i(E^i_{T,x},E^i_{T,y})$, where $E^i_{T,x}=E^i \sin \theta^i \cos \phi^i, E^i_{T,y}=E^i \sin \theta^i \sin \phi^i$. Here $E^i$ is the energy deposited in $i^{th}$ calorimeter cell and the angles $\theta^i$ and $\phi^i$ define the direction from the origin of the coordinate system to the $i^{th}$ calorimeter cell. We also define $\met \equiv |\vmet|$. 

\bibitem{alpgen}
M.~L.~Mangano {\sl et al.}, J.\ High Energy Phys.\ {\bf 07}, 001 (2003). We use version 2.05.

\bibitem{pythia6}
T.~Sj\"ostrand, S.~Mrenna and P.~Skands, J.\ High Energy Phys.\ {\bf 05}, 026 (2006). We use version
6.421 for generating the signal MC events.

\bibitem{CTEQ6L1}
J. Pumplin {\sl et al.}, J.\ High Energy Phys.\ {\bf 07}, 012 (2002); 
D.~Stump {\sl et al.}, J.\ High Energy Phys.\ {\bf 10}, 046 (2003).

\bibitem{MCFM}
J.~Campbell and R.~K.~Ellis, Phys.\ Rev.\ D {\bf 65}, 113007 (2002);
J.~Campbell, R.~K.~Ellis and D.~Rainwater, Phys.\ Rev.\ D {\bf 68}, 094021 (2003).

\bibitem{ttCrossSection}
S.~Moch and P.~Uwer, Phys.\ Rev.\ D {\bf 78}, 034003 (2008).

\bibitem{CTEQ5}
H.~L.~Lai {\it et al.} (CTEQ Collaboration), Eur.\ Phys.\ J.\ C {\bf 12}, 375 (2000).

\bibitem{geant}
R.~Brun, F.~Carminati and S.~Giani, CERN-W5013 (1993).

\bibitem{chargeflip}
V.~M.~Abazov {\it et al.} (D0 Collaboration), Phys.\ Rev.\ D {\bf 84}, 092002 (2011).

\bibitem{TMVA}
A.~Hoecker {\it et al.}, PoS {\bf ACAT}, 040 (2007).

\bibitem{bib-junk}
T. Junk, Nucl. Instrum. Methods in Phys. Res. A {\bf 434}, 435 (1999); 
A. Read, J. Phys. G {\bf 28}, 2693 (2002).

\bibitem{collie}
W.~Fisher, FERMILAB-TM-2386-E (2006).

\end{thebibliography}
\end{document}